\documentclass[preprint,12pt]{elsarticle}
\biboptions{sort&compress}

\usepackage[top=1in, bottom=1.25in, left=1.25in, right=1.25in]{geometry}
\usepackage{epsfig}         
\usepackage{cuted}
\usepackage{amsmath}
\usepackage{amssymb}
\usepackage{float}
\usepackage{xcolor}         
\usepackage[normalem]{ulem} 
\usepackage{multirow}
\usepackage{multicol}
\usepackage{subcaption}
\usepackage{verbatim}
\usepackage{MnSymbol}
\usepackage{psfrag}
\usepackage{bm}

\DeclareMathOperator{\sinc}{sinc}

\journal{Journal}

\begin{document}

\begin{frontmatter}

\title{Wave attenuation in viscoelastic hierarchical plates}

\author{Vin\'icius F. Dal Poggetto\corref{label1}\fnref{label1}}
\ead{v.fonsecadalpoggetto@unitn.it}

\author{Edson J. P. Miranda Jr.\fnref{label2,label3,label4}\corref{label1}}
\ead{edson.jansen@ifma.edu.br}

\author{Jos\'e Maria C. Dos Santos \fnref{label4}}

\author{Nicola M. Pugno\fnref{label1,label5}}

\fntext[label1]{Laboratory for Bio-inspired, Bionic, Nano, Meta Materials \& Mechanics, Department of Civil, Environmental and Mechanical Engineering, University of Trento, 38123 Trento, Italy}

\fntext[label2]{Federal Institute of Maranh\~{a}o, IFMA-EIB-DE, Rua Afonso Pena, 174, CEP 65010-030, S\~{a}o Lu\'{i}s, MA, Brazil}

\fntext[label3]{Federal Institute of Maranh\~{a}o, IFMA-PPGEM, Avenida Get\'{u}lio Vargas, 4, CEP 65030-005, S\~{a}o Lu\'{i}s, MA, Brazil}

\fntext[label4]{University of Campinas, UNICAMP-FEM-DMC, Rua Mendeleyev, 200, CEP 13083-970, Campinas, SP, Brazil}

\fntext[label5]{School of Engineering and Materials Science, Queen Mary University of London, Mile End Road, London E1 4NS, United Kingdom}

\cortext[label1]{Corresponding authors}


\begin{abstract}
{\it 
Phononic crystals (PCs) are periodic structures obtained by the spatial arrangement of materials with contrasting properties, which can be designed to efficiently manipulate mechanical waves. Plate structures can be modeled using the Mindlin-Reissner plate theory and have been extensively used to analyze the dispersion relations of PCs. Although the analysis of the propagating characteristics of PCs may be sufficient for simple elastic structures, analyzing the evanescent wave behavior becomes fundamental if the PC contains viscoelastic components. Another complication is that increasingly intricate material distributions in the unit cell of PCs with hierarchical configuration may render the calculation of the complex band structure (i.e., considering both propagating and evanescent waves) prohibitive due to excessive computational workload.
In this work, we propose a new extended plane wave expansion formulation to compute the complex band structure of thick PC plates with arbitrary material distribution using the Mindlin-Reissner plate theory containing constituents with a viscoelastic behavior approximated by a Kelvin-Voigt model.
We apply the method to investigate the evanescent behavior of periodic hierarchically structured plates for either (i) a hard purely elastic matrix with soft viscoelastic inclusions or (ii) a soft viscoelastic matrix with hard purely elastic inclusions. Our results show that for (i), an increase in the hierarchical order leads to a weight reduction with relatively preserved attenuation characteristics, including attenuation peaks due to locally resonant modes that present a decrease in attenuation upon increasing viscosity levels. For (ii), changing the hierarchical order implies in opening band gaps in distinct frequency ranges, with an overall attenuation improved by an increase in the viscosity levels.
}
\end{abstract}




\begin{keyword}
Plane wave expansion \sep Hierarchical structure \sep Mindlin-Reissner plate \sep Evanescent wave \sep Viscoelastic material.
\end{keyword}


\end{frontmatter}

\section{Introduction} \label{introduction}

Hierarchical materials present a structured composition across multiple length scales and have long been a subject of study \cite{lakes1993materials} since their occurrence in nature is associated with excellent static \cite{meyers2013structural} and dynamic characteristics \cite{bosia2022optimized}. Such properties may be harnessed to design novel materials through the selection of constituents in a proper multi-level structuring \cite{chen2021advances}. Although the static properties of hierarchical structures have been thoroughly exposed in different contexts \cite{chen2016hierarchical,meza2015resilient,kochmann2019multiscale}, their application to obtain interesting dynamic properties remains to be fully explored. In particular, hierarchical periodic structures can be used to attenuate waves in a particularly broad manner \cite{zhang2013broadband,chen2015multiband,poggetto2021band}, which has been demonstrated both theoretically and experimentally with the use of dissipative elastic metamaterials \cite{miniaci2018design}.

Periodic structures, which can be obtained by the repetition of a representative unit cell, are known for their ability to manipulate waves \cite{brillouin1953wave}, leading to applications in mechanical systems such as vibration attenuation, imaging, and cloaking \cite{craster2012acoustic,deymier2013acoustic}. A remarkable feature that can be found in a specific class of periodic structures named phononic crystals (PCs) \cite{khelif2015phononic} is that impedance mismatches achieved, for instance, by using spatial modulations in single-phase materials \cite{poggetto2020widening,bibi2019manipulation,pelat2019control,sorokin2016effects} or by combining materials with contrasting elastic properties \cite{kushwaha1994theory,kushwaha1994band,wilm2003out} can lead to the occurrence of frequency ranges named band gaps (BGs). Such frequency ranges are typically created in PCs by the destructive interference of waves (Bragg scattering) \cite{romero2019fundamentals}, thus prohibiting free wave propagation due to the resulting purely evanescent behavior of waves \cite{krushynska2017coupling,laude2009evanescent}.
In the case of locally resonant PCs \cite{liu2000locally}, Fano-like interference mechanisms can also occur in the sub-wavelength scale \cite{goffaux2002evidence}, thus typically leading to low-frequency BGs.
Although the opening of BGs is evident in the case of purely elastic materials, the inclusion of damping leads to complications in the determination of BGs, since spatial attenuation becomes inherent in such cases \cite{laude2013effect,krushynska2016visco,van2017impact}.

Band diagrams can be used to conveniently analyze the dispersion relation (i.e., the relation between the wavenumber and frequency) of periodic structures and can be obtained through a variety of techniques. Although finite element (FE)-based techniques are widely employed \cite{duhamel2006finite,mace2008modelling,mencik2010low,nobrega2016vibration}, these methods usually suffer from disadvantages in terms of computational burden, which may render their use prohibitive when distinct orders of hierarchy are considered due to the inherently detailed modeling which is required. Dispersion relations can also be computed using the plane wave expansion (PWE) method, which typically results in a reduced computational effort \cite{bin2011improved,beli2018wave,poggetto2020flexural}. On the other hand, the PWE method usually requires the use of analytical expressions for the shape of the scatterers included in the PC matrix material, which limits the applicability of the method. Also, the conventional PWE method does not offer information about the evanescent behavior of waves \cite{laude2009evanescent,romero2010evidences}, which is necessary to characterize the complex band structure of damped systems. A solution to this limitation is proposed by the extended plane wave expansion (EPWE) method \cite{hsue2005extended}, which yields both propagating and evanescent parts of the wave vectors for a given frequency of interest at the expense of a greater computational cost.

Recent advances in the experimental observations of guided waves in biological structures such as the human skull \cite{estrada2018observation} have revealed the propagation of Lamb waves \cite{estrada2018looking,mazzotti2021experimental}, which motivates the analysis of the wave propagation in structured media using plate theories \cite{gao2019free}.
Plates have been thoroughly explored as versatile structures in the field of metamaterials and PCs using the Kirchhoff plate theory \cite{ventsel2001thin} with periodic arrays of embedded resonators \cite{xiao2008flexural,gao2019single},
periodic arrays of local resonators \cite{xiao2012flexural,miranda2019flexural},
the inclusion of point defects \cite{yao2009propagation},
or the Mindlin plate theory \cite{timoshenko1959theory} with embedded \cite{hsu2010plate,poggetto2020flexural} or attached resonators \cite{miranda2020wave,miranda2022plane}. The use of plate theories for the computation of dispersion relations seems thus to be the most common solution when compared to the use of solid models with stress-free boundaries \cite{tanaka1998surface} or equivalent low impedance surrounding media \cite{vasseur2008absolute}.
Although the Kirchhoff-Love plate theory can be considered under the assumption of negligible shear strain and rotational inertia in the low-frequency range, its use may require additional refinements of the kinematic model or adjustments to properly include inertial terms \cite{kaplunov1998dynamics,goldenveizer1993timoshenko}. On the other hand, the Mindlin-Reissner plate theory already accounts for shear strain and rotational inertia terms, although requiring larger computational models, being also more suited to analyze structures which operate in higher frequency ranges, which is the case of PCs. Previous works have computed the dispersion relation considering the viscoelastic material behavior for the SH-wave of two-dimensional PCs \cite{li2021analysis} and quasi-periodic lattices \cite{mukhopadhyay2019frequency}. The investigation of the effects of hierarchical structuring on plates, however, especially when considering the complex band structure necessary to fully understand the implications of components that present damping, remains largely unexplored.

In this work, we propose the numerical investigation of the evanescent behavior of viscoelastic hierarchical plate PCs with the use of the EPWE method applied using the Mindlin-Reissner plate theory. This paper is organized as follows: Section \ref{models_methods} presents the considered plate theory, the material behavior, and the application of the EPWE method to plates with discrete geometries. Section \ref{results} presents the obtained results, and Section \ref{conclusions} presents our concluding remarks.


\section{Models and methods} \label{models_methods}

In this section, we present the analytical derivations regarding the calculation of dispersion curves for periodic PC structures using the Mindlin-Reissner plate theory, which considers non-negligible rotational inertia and transverse shear strain \cite{liew1995research}. The related dynamic equations are expressed considering periodic solutions for both displacements and rotations, and also periodic material properties. Then, a Kelvin-Voigt viscoelastic model is included to represent the dissipative behavior of constituents, which is considered when formulating the equations that allow to compute the complex band structure of the unit cell.

\subsection{Wave propagation in periodic plates using the Mindlin-Reissner plate theory}

Plate theories can be employed for the analysis of structural elements with one dimension (thickness) considerably smaller than the other two ones \cite{leissa1969vibration}.
Considering the Mindlin-Reissner plate theory, the equation that describes the dynamic behavior in the time domain ($t$) of an isotropic plate lying in the $xy$ plane without applied loads is given by
\begin{subequations} \label{plate_mindlin}
\begin{align}
 \frac{\partial}{\partial x} \bigg[ \kappa \mu h \bigg( \frac{\partial u_z}{\partial x} - \psi_x \bigg) \bigg] +
 \frac{\partial}{\partial y} \bigg[ \kappa \mu h \bigg( \frac{\partial u_z}{\partial y} - \psi_y \bigg) \bigg] &= 
 \rho h \frac{\partial^2 u_z}{\partial t^2} \, , \label{dynamic_plate_mindlin_a} \\
 \frac{\partial}{\partial x} \bigg[ D \bigg( \frac{\partial \psi_x}{\partial x} + \nu \frac{\partial \psi_y}{\partial y} \bigg) \bigg]  + 
 \frac{\partial}{\partial y} \bigg[ \frac{D(1-\nu)}{2} \bigg( \frac{\partial \psi_x}{\partial y} + \frac{\partial \psi_y}{\partial x} \bigg) \bigg] +
 \kappa \mu h \bigg( \frac{\partial u_z}{\partial x} - \psi_x \bigg) &=
 \frac{\rho h^3}{12} \frac{\partial ^2 \psi_x}{\partial t^2} \, , \label{dynamic_plate_mindlin_b} \\
 \frac{\partial}{\partial x} \bigg[ \frac{D(1-\nu)}{2} \bigg( \frac{\partial \psi_x}{\partial y} + \frac{\partial \psi_y}{\partial x} \bigg) \bigg]  + 
 \frac{\partial}{\partial y} \bigg[ D \bigg( \nu \frac{\partial \psi_x}{\partial x} + \frac{\partial \psi_y}{\partial y} \bigg) \bigg] + 
 \kappa \mu h \bigg( \frac{\partial u_z}{\partial y} - \psi_y \bigg) &=
 \frac{\rho h^3}{12} \frac{\partial ^2 \psi_y}{\partial t^2} \, , \label{dynamic_plate_mindlin_c}
\end{align}
\end{subequations}
where $u_z = u_z(x,y,t)$ is the plate out-of-plane displacement, $\psi_x = \psi_x(x,y,t)$ and $\psi_y = \psi_x(x,y,t)$ represent the rotations of the plate midplane normal direction, $\kappa$ is the plate shear correction factor \cite{cook2001concepts}, $\mu=\mu(x,y)$ is the material shear modulus, $\nu=\nu(x,y)$ is the material Poisson's ratio, $\rho=\rho(x,y)$ is the material mass density, $h$ is the plate thickness, and $D=D(x,y)$ is the plate flexural stiffness, given by
\begin{equation} \label{plate_flexural}
  D(x,y) = \frac{E(x,y) \, h^3}{12(1-\nu^2(x,y))} \, ,
\end{equation}
where $E=E(x,y)$ is the material Young's modulus. If the plate material properties are periodic, the resulting displacement and rotation solutions present the same periodicity \cite{bloch1929quantenmechanik}, which can be used to obtain the dispersion curves of the periodic medium.


To properly analyze the propagation of elastic waves in periodic plates, the PWE method requires the expression of displacements, rotations, and material properties considering their respective spatial periodicity. Let us denote the position vector $\mathbf{r}$ in terms of its Cartesian components, i.e., $\mathbf{r} = x \hat{\mathbf{i}} + y \hat{\mathbf{j}}$. Considering a Bloch solution \cite{bloch1929quantenmechanik} for the displacements $u_z(x,y,t) = u_z(\mathbf{r},t)$, one has
\begin{equation} \label{disp1}
  u_z(\mathbf{r},t) = u_{z \mathbf{k}}(\mathbf{r}) e^{-\text{i}\omega t} \, ,
\end{equation}
where $u_{z \mathbf{k}}(\mathbf{r})$ is a spatial function, and $\omega$ is the considered angular frequency. According to Bloch's theorem, $u_{z \mathbf{k}}(\mathbf{r})$ must obey
\begin{equation} \label{disp2}
  u_{z \mathbf{k}}(\mathbf{r}) = e^{\text{i} \mathbf{k} \cdot \mathbf{r} } \, u_{z0}(\mathbf{r}) \, ,
\end{equation}
where $\mathbf{k}$ is the two-dimensional wave vector, which can be written in terms of its Cartesian components as $\mathbf{k} = k_x \hat{\mathbf{i}} + k_y \hat{\mathbf{j}}$, and $u_{z0}(\mathbf{r})$ is a periodic function with the same periodicity as the medium. Thus, $u_{z0}(\mathbf{r})$ can be represented as a Fourier series using
\begin{equation} \label{disp3}
 u_{z0}(\mathbf{r}) = \sum_{\mathbf{G}} \hat{u}_z(\mathbf{G}) e^{\text{i} \mathbf{G} \cdot \mathbf{r}} \, ,
\end{equation}
where $\hat{u}_z(\mathbf{G})$ denotes a Fourier coefficient for the representation of the out-of-plane displacements, which must be summed for infinite reciprocal lattice vectors of the form $\mathbf{G} = G_x \hat{\mathbf{i}} + G_y \hat{\mathbf{j}}$. For a square lattice unit cell of side length $a$, $G_x = n_x \frac{2\pi}{a}$ and $G_y = n_y \frac{2\pi}{a}$ for $\{ n_x, \, n_y \} \in \mathbb{Z}$. Thus, for $-N \leq \{ n_x, \, n_y \} \leq N$, $N \in \mathbb{N}$, a total of $n_G = (1 + 2 N)^2$ plane waves is considered.

The expression of displacements in the periodic medium can thus be obtained by combining Eqs. (\ref{disp1})-(\ref{disp3}) in the form
\begin{equation} \label{disp_final}
  u_z(\mathbf{r},t) = e^{-\text{i}\omega t} \sum_{\mathbf{G}} \hat{u}_z(\mathbf{G}) e^{\text{i} (\mathbf{k} + \mathbf{G}) \cdot \mathbf{r}} \, ,
\end{equation}
which presents a form appropriate for its application in the PWE method.

An analogous procedure can be applied to the rotations $\psi_x(\mathbf{r},t)$ and $\psi_y(\mathbf{r},t)$, allowing to write
\begin{equation} \label{rotations_final}
\begin{aligned}
  \psi_x(\mathbf{r},t) = e^{-\text{i}\omega t} \sum_{\mathbf{G}} \hat{\psi}_x(\mathbf{G}) e^{\text{i} (\mathbf{k} + \mathbf{G}) \cdot \mathbf{r}} \, , \\
  \psi_y(\mathbf{r},t) = e^{-\text{i}\omega t} \sum_{\mathbf{G}} \hat{\psi}_y(\mathbf{G}) e^{\text{i} (\mathbf{k} + \mathbf{G}) \cdot \mathbf{r}} \, ,
\end{aligned}
\end{equation}
where $\hat{\psi}_x(\mathbf{G})$ and $\hat{\psi}_y(\mathbf{G})$ denote the Fourier coefficients for the representation of the midplane rotations $\psi_x$ and $\psi_y$, respectively.

The wave vector $\mathbf{k}$ and its components can be classified according to its real and imaginary parts, assumed to be co-linear \cite{collet2011floquet}: purely real wave vectors yield propagating waves, purely imaginary wave vectors yield evanescent waves, and complex conjugate solutions yield decaying propagating waves.

\subsection{Material properties}

\begin{figure}[h!]
  \centering
  \includegraphics[height=6cm]{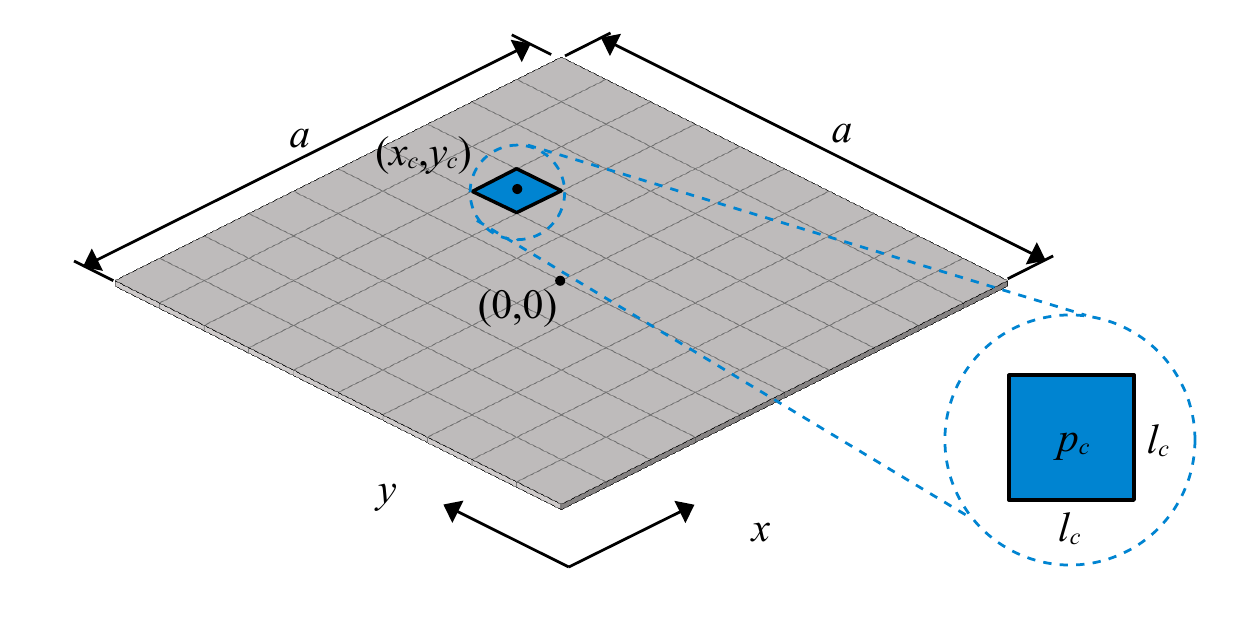}
  \caption{Square plate with side length $L$, uniformly divided in pixels with constant property $p_c$ for each corresponding domain $x \in ( x_c - l_c/2 , \, x_c + l_c/2)$, $y \in (y_c - l_c/2, \, y_c + l_c/2)$.}
  \label{plate_pixels}
\end{figure}

Let us consider a square unit cell with a side length of $a$, divided in a set of square elements (pixels) used to describe its spatial configuration. Each pixel can be described by the $x$- and $y$-coordinates of its center, denoted as $x_c$ and $y_c$, respectively, its side length $l_c$, and a corresponding material property $p_c$ (Figure \ref{plate_pixels}). Although the pixels do not necessarily form a regular grid, they cover the entire area $A$ of the unit cell, i.e., $\sum_A l_c^2 = a^2$. Furthermore, a four-fold symmetry is assumed so that the band structure of the medium can be investigated by analyzing a reduced region of the unit cell \cite{maurin2018probability}.

The PWE method also requires that the material properties be written in terms of their Fourier series. Thus, a general expression can be stated for a given material property $p$ using
\begin{equation} \label{p_fourier}
 p(\mathbf{r}) = \sum_{\mathbf{G}} \, \hat{p}(\mathbf{G}) e^{\text{i} \mathbf{G} \cdot \mathbf{r}} \, ,
\end{equation}
where $\hat{p}(\mathbf{G})$ denotes a Fourier series term for the representation of the corresponding material property, which theoretically must be summed for an infinite number of reciprocal lattice vectors $\mathbf{G}$.

By considering the description of the unit cell as an ensemble of square pixels, as previously described, the term $\hat{p}(\mathbf{G})$ can be computed as \cite{zarei2008symmetry,liu2014band}
\begin{equation} \label{phat_G}
 \hat{p}(\mathbf{G}) = \sum_{A}
 \bigg( \frac{l_c}{a} \bigg)^2 \,
 \sinc \bigg( \frac{G_x l_c}{2} \bigg) \sinc \bigg( \frac{G_y l_c}{2} \bigg)
 \,
 p_c \, e^{-\text{i} G_x x_c} e^{-\text{i} G_y y_c}
  \, ,
\end{equation}
which must be summed over the whole unit cell, i.e., for $\{ x_c \pm l_c/2 , \, y_c \pm l_c/2 \} \in [-a/2, \, a/2]$.

\begin{figure}[h!]
  \centering
  \includegraphics[width=15cm]{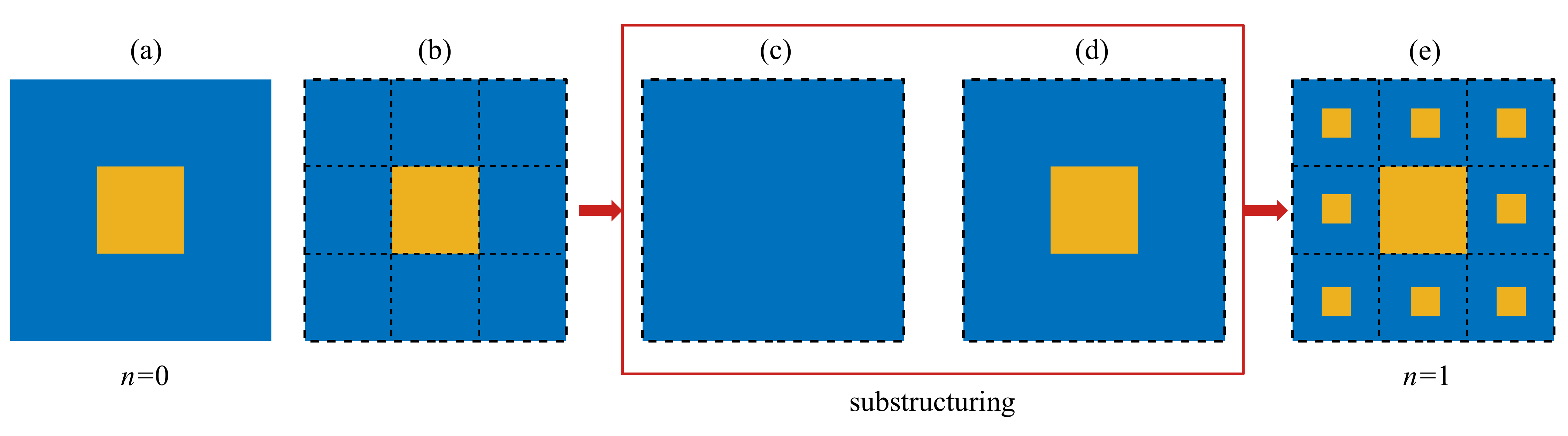}
  \caption{Hierarchical plate structuring.
  (a) The initial unit cell, with a $1/9$ filling fraction of the inclusion material (yellow) over the matrix material (blue), (b) is further divided in a $3 \times 3$ regular grid, (c) where each matrix pixel is (d) substituted by a scaled version of the initial unit cell, while for inclusion pixels, no substitutions are made.
  (e) The resulting self-similar structure represents a hierarchical structure of a higher order.}
  \label{plate_structuring}
\end{figure}

The process of construction of the hierarchical structure starts from a square unit cell composed of a matrix material with the inclusion of another material corresponding to a $1/9$ area filling fraction (shown in blue and yellow, respectively, in Figure \ref{plate_structuring}a). This initial structure is further divided in a $3 \times 3$ grid (dashed lines in Figure \ref{plate_structuring}b), where each matrix pixel is substituted by a scaled version of the initial unit cell (Figures \ref{plate_structuring}c and \ref{plate_structuring}d), while for inclusion pixels no substitutions are made. This procedure is applied to each of the $9$ substructures, thus obtaining a self-similar Sierpinski carpet fractal structure \cite{sierpinski1916courbe} with an increased order (Figure \ref{plate_structuring}e) \cite{zhang2016analyzing,huang2017multiple}. This process can be repeated indefinitely.
The first presented structure is here considered as the $0$-th order hierarchical structure, since the inclusion and matrix materials present the same orders of magnitude.
It is also important to note that the resulting unit cell presents a structure which is especially suited to be used in combination with Eq. (\ref{phat_G}), thus allowing to compute the corresponding representation of the periodic material properties.

The inclusion filling fraction for the $n$-th hierarchical order, $\phi_n$, is given by
\begin{equation} \label{filling_fraction}
 \phi_n = 1 - \bigg( \frac{8}{9} \bigg)^{n+1}, \,
\end{equation}
where we consider $n=0$ as the first order. The equivalent specific mass density is given by
\begin{equation} \label{equivalent_rho}
 \overline{\rho}_n =  \rho_m \bigg( \frac{8}{9} \bigg)^{n+1} +
  \rho_i \bigg[ 1 - \bigg( \frac{8}{9} \bigg)^{n+1} \bigg] \, ,
\end{equation}
where $\rho_m$ and $\rho_i$ refer, respectively, to the matrix and inclusion material mass densities.

\subsection{Viscoelastic material behavior}

The real part of wave vectors shown in band diagrams are commonly computed considering a purely elastic material behavior. However, viscoelastic behavior is observed in several components typically used in mechanical systems \cite{ferry1980viscoelastic}. The frequency dependence of the material shear modulus can be approximated by simple models analogous to spring and dashpot elements \cite{lakes2009viscoelastic}. The Kelvin-Voigt viscoelastic model can be used to describe the material dissipation proportional to the excitation, relating the shear stress, $\tau(\mathbf{r},t)$, and the shear strain, $\gamma(\mathbf{r},t)$, through the time-dependent relation
\begin{equation}
 \tau(\mathbf{r},t) = \mu'(\mathbf{r}) \gamma(\mathbf{r},t) + \mu''(\mathbf{r}) \frac{\partial \gamma(\mathbf{r},t)}{\partial t} \, ,
\end{equation}
where $\mu'(\mathbf{r})$ is the shear storage modulus and $\mu''(\mathbf{r})$ represents a velocity-proportional viscoelastic dissipative term. When assuming the single-frequency time functions $\tau(\mathbf{r},t) = e^{-\text{i}\omega t} \tau(\mathbf{r},\omega)$ and $\gamma(\mathbf{r},t) = e^{-\text{i}\omega t} \gamma(\mathbf{r},\omega)$ this equation can be represented in the frequency domain, as
\begin{equation} \label{kelvin_voigt}
 \tau(\mathbf{r},\omega) = \mu^*(\mathbf{r},\omega) \gamma(\mathbf{r},\omega) \, ,
\end{equation}
where $\mu^*(\mathbf{r},\omega) = \mu'(\mathbf{r}) - \text{i} \omega \mu''(\mathbf{r})$ is the complex shear modulus of the Kelvin-Voigt material; henceforth, the $(\cdot)^*$ superscript is used to denote a viscoelastic quantity. The imaginary part of the complex shear modulus is also often referred to as shear loss modulus. It is important to notice that the negative sign in the imaginary part of $\mu^*(\mathbf{r},\omega)$ is a consequence of assuming time functions of the form $e^{-\text{i}\omega t}$, i.e., with the same form as the displacement function given by Eq. (\ref{disp1}).

For simplification, let us assume that the Poisson's ratio of the structure materials have a constant value over the considered frequency range. Although the frequency dependence of the Poisson's ratio may present a rather complex behavior, which is often ignored \cite{tschoegl2002poisson}, this simplifying hypothesis seems to yield reasonable experimental results \cite{merheb2008elastic,krushynska2021dissipative}. Thus, this allows to write the Young's modulus of a given material as
\begin{equation}
 E(\mathbf{r},t) = 2(1+\nu(\mathbf{r})) \, \mu(\mathbf{r},t)  \, ,
\end{equation}
which yields the frequency domain relation
\begin{equation}
 E^*(\mathbf{r},\omega) = E'(\mathbf{r}) - \text{i} \omega E''(\mathbf{r})
 \, ,
\end{equation}
where $E'(\mathbf{r}) = 2(1+\nu(\mathbf{r})) \, \mu'(\mathbf{r})$ and $E''(\mathbf{r}) = 2(1+\nu(\mathbf{r})) \, \mu''(\mathbf{r})$. Consequently,
the spatial- and frequency-dependent plate flexural stiffness, $D^*(\mathbf{r},\omega)$, can be computed as
\begin{equation} \label{D_ve_rw}
 D^*(\mathbf{r},\omega) = \frac{E^*(\mathbf{r},\omega) h^3}{12(1-\nu^2(\mathbf{r}))} = D'(\mathbf{r}) - \text{i} \omega D''(\mathbf{r}) \, ,
\end{equation}
where $D'(\mathbf{r}) = \frac{E'(\mathbf{r}) h^3}{12(1-\nu^2(\mathbf{r}))}$, $D''(\mathbf{r}) = \frac{E''(\mathbf{r}) h^3}{12(1-\nu^2(\mathbf{r}))}$. We also define the auxiliary quantities $\alpha^*(\mathbf{r},\omega)$ and $\beta^*(\mathbf{r},\omega)$, which will be present in the PWE derivations, as
\begin{equation} \label{alpha_beta_ve_rw}
\begin{aligned}
 \alpha^*(\mathbf{r},\omega) &= 
 D^*(\mathbf{r},\omega) \nu(\mathbf{r}) =
 \alpha'(\mathbf{r}) - \text{i} \omega \alpha''(\mathbf{r})
 \, , \\
 \beta^*(\mathbf{r},\omega) &=
 \frac{ D^*(\mathbf{r},\omega) (1-\nu(\mathbf{r})) }{2} =
 \beta'(\mathbf{r}) - \text{i} \omega \beta''(\mathbf{r}) \, ,
\end{aligned}
\end{equation}
where
$\alpha'(\mathbf{r}) = \frac{E'(\mathbf{r}) h^3 \nu(\mathbf{r})}{12(1-\nu^2(\mathbf{r}))}$,
$\alpha''(\mathbf{r}) = \frac{E''(\mathbf{r}) h^3 \nu(\mathbf{r})}{12(1-\nu^2(\mathbf{r}))}$,
$\beta'(\mathbf{r}) = \frac{E'(\mathbf{r}) h^3 }{24(1+\nu(\mathbf{r}))}$, and
$\beta''(\mathbf{r}) = \frac{E''(\mathbf{r}) h^3 }{24(1+\nu(\mathbf{r}))}$.

For purely elastic materials, some wave vectors can be expected to be purely real (propagating waves). However, if one also considers the damping owing to viscoelastic effects, all waves can be expected to present some degree of decay, and the analysis of the real part of wavenumbers is no longer sufficient to completely describe the computed band diagrams. In these cases, the EPWE method can be employed to obtain the evanescent behavior of waves.

\subsection{Extended plane wave expansion method}

The Fourier representations of spatial- and frequency-dependent material properties
$\mu^*(\mathbf{r},\omega)$ (see Eq. (\ref{kelvin_voigt})), $D^*(\mathbf{r},\omega)$ (see Eq. (\ref{D_ve_rw})), $\alpha^*(\mathbf{r},\omega)$, $\beta^*(\mathbf{r},\omega)$ (see Eq. (\ref{alpha_beta_ve_rw})), and $\rho(\mathbf{r})$ can be expressed using Eq. (\ref{p_fourier}) and summarized as
\begin{subequations} \label{materials_fourier}
\begin{align}
 p^*(\mathbf{r},\omega) &=
 \sum_{\mathbf{G}}
 \hat{p}^*(\mathbf{G}) e^{\text{i} \mathbf{G} \cdot \mathbf{r}}
 =
 \sum_{\mathbf{G}}
 ( \hat{p}'(\mathbf{G}) - \text{i} \omega \hat{p}''(\mathbf{G}) )
 e^{\text{i} \mathbf{G} \cdot \mathbf{r}} \, , \label{materials_fourier_p} \\
 \rho(\mathbf{r}) &= \sum_{\mathbf{G}} \hat{\rho}(\mathbf{G}) e^{\text{i} \mathbf{G} \cdot \mathbf{r}} \, , \label{materials_fourier_rho_r}
\end{align}
\end{subequations}
where $p^*(\mathbf{r},\omega)$ may refer to $\mu^*(\mathbf{r},\omega)$, $D^*(\mathbf{r},\omega)$, $\alpha^*(\mathbf{r},\omega)$, or $\beta^*(\mathbf{r},\omega)$; $\hat{p}'(\mathbf{G})$, $\hat{p}''(\mathbf{G})$, and $\hat{\rho}(\mathbf{G})$ denote the Fourier components for the spatial expressions of $p'(\mathbf{r})$, $p''(\mathbf{r})$, and $\rho(\mathbf{r})$, respectively.

Substituting the out-of-plane displacement (Eq. (\ref{disp_final})), rotations (Eq. (\ref{rotations_final})), and material expressions (Eq. (\ref{materials_fourier})) in the plate first dynamic equation (Eq. (\ref{dynamic_plate_mindlin_a})), one may write 
\begin{equation} \small
\begin{aligned} 
 \sum_{\mathbf{H}} \sum_{\mathbf{G}}
 & \bigg\{
 [ \kappa \hat{\mu}^*(\mathbf{H}) h ((k_x+G_x)(k_x+G_x+H_x) + (k_y+G_y)(k_y+G_y+H_y)) - \omega^2 \hat{\rho}(\mathbf{H}) h ] \hat{u}_z (\mathbf{G}) \\
 &+ \text{i} \, \kappa \hat{\mu}^*(\mathbf{H}) h (k_x+G_x+H_x) \hat{\psi}_x(\mathbf{G})
  + \text{i} \, \kappa \hat{\mu}^*(\mathbf{H}) h (k_y+G_y+H_y) \hat{\psi}_y(\mathbf{G})
 \bigg\} e^{\text{i}(\mathbf{k}+\mathbf{G}+\mathbf{H})\cdot \mathbf{r}} = 0 \, .
\end{aligned}
\end{equation}

The orthogonality property of the complex exponential \cite{hsu1967fourier,strang1988linear} can be used to obtain, substituting $\mathbf{H} = \mathbf{G}_i - \mathbf{G}_j$ and $\mathbf{G} = \mathbf{G}_j$, for every $\mathbf{r}$, the set of linear equations $i = 1, \, \ldots, \, n_G$ written as
\begin{equation} \label{mindlin_pwe_uz}
\small
\begin{aligned} 
 \sum_{\mathbf{G}_j}
 & [ \kappa \hat{\mu}_{ij}^* h 
 ((k_x+G_{xj})(k_x+G_{xi}) + (k_y+G_{yj})(k_y+G_{yi})) - \omega^2 \hat{\rho}_{ij} h ]
 \hat{u}_z (\mathbf{G}_j) \\
 &+ \text{i} \, \kappa \hat{\mu}_{ij}^* h (k_x+G_{xi}) \hat{\psi}_x(\mathbf{G}_j)
 + \text{i} \, \kappa \hat{\mu}_{ij}^* h (k_y+G_{yi}) \hat{\psi}_y(\mathbf{G}_j)
 = 0 \, ,
\end{aligned}
\end{equation}
where $\hat{\mu}_{ij}^* = \hat{\mu}'(\mathbf{G}_i-\mathbf{G}_j) - \text{i} \, \omega \hat{\mu}''(\mathbf{G}_i-\mathbf{G}_j)$ and
$ \hat{\rho}_{ij} = \hat{\rho}(\mathbf{G}_i-\mathbf{G}_j)$. In this case, each material property $p(\mathbf{r})$ represented in terms of its Fourier series (i.e., $\mu'$, $\mu''$, and $\rho$) can be associated with a symmetric matrix $\mathbf{p}$, with terms given by $[\mathbf{p}]_{ij} = p(\mathbf{G}_i - \mathbf{G}_j)$, which can be computed with the use of Eq. (\ref{phat_G}) for $\mathbf{G} = \mathbf{G}_i-\mathbf{G}_j$.

It is important to mention a particularity associated with the use of Fourier series for the analysis of discontinuous structures \cite{li1996use}. In most usual cases, the terms $h_j = \sum_j f_{i-j} g_j$ present in the last equation (e.g., $ \sum_{\mathbf{G}_j} \hat{\mu}'(\mathbf{G}_i-\mathbf{G}_j) \hat{u}_z(\mathbf{G}_j)$) are in the form of Laurent's rule.
This is commonly used to justify the formulation named improved plane wave expansion (IPWE) \cite{cao2004convergence} to improve the convergence of the PWE method, formulated by substituting the previous sum by the inverse rule $h_j = \sum_j [1/f]^{-1}_{i-j} g_j$, where $[f]^{-1}$ denotes the inverse of the Toeplitz matrix $[f]_{ij} = f_{i-j}$.
In the present case, the use of Eq. (\ref{phat_G}) to compute the Fourier expansion of material properties considering $p(\mathbf{G}_i-\mathbf{G}_j)$ does not lead to a Toeplitz matrix, which may possibly hinder the improvement yielded by the IPWE. Also, since some of the components may be purely elastic and thus present $p''=0$, the IPWE method is not applicable for the computation of all matrices, being restricted to the matrices associated with $\mu'$, $D'$, $\alpha'$, and $\beta'$.

The same reasoning use to obtain Eq. (\ref{mindlin_pwe_uz}) from Eq. (\ref{dynamic_plate_mindlin_a}) can be applied to the plate dynamic equation given by Eq. (\ref{dynamic_plate_mindlin_b}) to obtain
\begin{equation} \label{mindlin_pwe_psix}
\small
\begin{aligned} 
 & \sum_{\mathbf{G}_j}
 \big[ \hat{D}^*_{ij}(k_x+G_{xj})(k_x+G_{xi}) +
 \hat{\beta}^*_{ij}(k_y+G_{yj})(k_y+G_{yi}) +
 \kappa \hat{\mu}^*_{ij} h - \omega^2 \hat{\rho}_{ij} h^3/12
 \big] \hat{\psi}_x(\mathbf{G}_j) + \\
 & \big[ \hat{\alpha}^*_{ij}(k_y+G_{yj})(k_x+G_{xi}) +
 \hat{\beta}^*_{ij}(k_x+G_{xj})(k_y+G_{yi}) \big] \hat{\psi}_y(\mathbf{G}_j)
 - \text{i} \, \kappa \hat{\mu}^*_{ij} h (k_x+G_{xj}) \hat{u}_z (\mathbf{G}_j) =0 \, ,
\end{aligned}
\end{equation}
where $\hat{D}^*_{ij} = D'(\mathbf{G}_i-\mathbf{G}_j) - \text{i} \, \omega D''(\mathbf{G}_i-\mathbf{G}_j)$, $\hat{\alpha}^*_{ij} = \alpha'(\mathbf{G}_i-\mathbf{G}_j) - \text{i} \, \omega \alpha''(\mathbf{G}_i-\mathbf{G}_j)$, and $\hat{\beta}^*_{ij} = \beta'(\mathbf{G}_i-\mathbf{G}_j) - \text{i} \, \omega \beta''(\mathbf{G}_i-\mathbf{G}_j)$, computed with the use of Eq. (\ref{phat_G}).

Analogously, the same procedure can be applied to Eq. (\ref{dynamic_plate_mindlin_c}), allowing to write
\begin{equation} \label{mindlin_pwe_psiy}
\small
\begin{aligned} 
 & \sum_{\mathbf{G}_j}
 \big[
 \hat{D}^*_{ij}(k_y+G_{yj})(k_y+G_{yi}) +
 \hat{\beta}^*_{ij}(k_x+G_{xj})(k_x+G_{xi}) +
 \kappa \hat{\mu}^*_{ij} h - \omega^2 \hat{\rho}_{ij} h^3/12
 \big] \hat{\psi}_y(\mathbf{G}_j) + \\
 & \big[ \hat{\alpha}^*_{ij}(k_x+G_{xj})(k_y+G_{yi}) +
 \hat{\beta}^*_{ij}(k_y+G_{yj})(k_x+G_{xi}) \big] \hat{\psi}_x(\mathbf{G}_j)
 - \text{i} \, \kappa \hat{\mu}^*_{ij} h (k_y+G_{yj}) \hat{u}_z (\mathbf{G}_j) =0 \, .
\end{aligned}
\end{equation}

By writing the Cartesian components of the wave vector in terms of the azimuth angle $\phi$, i.e., $k_x = k \cos \phi$ and $k_y = k \sin \phi$, Eqs. (\ref{mindlin_pwe_uz})--(\ref{mindlin_pwe_psiy}) can be used, considering a finite set of reciprocal lattice vectors $\mathbf{G}_1$, $\mathbf{G}_2$, $\ldots$, $\mathbf{G}_{n_G}$, to obtain a set of equations that can be formulated as the polynomial eigenproblem
\begin{equation} \label{poly_eigen_kw}
 ( k^2 \mathbf{A}_2(\omega,\varphi) + k \mathbf{A}_1(\omega,\varphi) + \mathbf{A}_0(\omega,\varphi) - \omega^2 \mathbf{B} ) \mathbf{V} = \mathbf{0} \, ,
\end{equation}
where matrices $\mathbf{A}_2(\omega,\varphi)$, $\mathbf{A}_1(\omega,\varphi)$, $\mathbf{A}_0(\omega,\varphi)$, and $\mathbf{B}$ are described in \ref{eigenproblem_matrices}.

Equation (\ref{poly_eigen_kw}) can be solved by rearranging it as a second-order polynomial eigenvalue problem in a companion matrix form \cite{edelman1995polynomial}, leading to the $k = k(\omega)$ formulation stated as
\begin{equation} \label{poly_eigen_kw_matrix}
 \left[
 \begin{array}{cc}
  \mathbf{0} & \mathbf{I}  \\
  \mathbf{A}_0(\omega,\varphi) - \omega^2 \mathbf{B} & \mathbf{A}_1(\omega,\varphi) \\
 \end{array}
 \right]
 \left\{
 \begin{array}{c}
  \mathbf{V} \\ k\mathbf{V}
 \end{array}
 \right\} = 
 k
 \left[
 \begin{array}{cc}
  \mathbf{I} & \mathbf{0}  \\
  \mathbf{0} & -\mathbf{A}_2(\omega,\varphi) \\
 \end{array}
 \right]
 \left\{
 \begin{array}{c}
  \mathbf{V} \\ k\mathbf{V}
 \end{array}
 \right\} \, ,
\end{equation}
which can be solved to obtain complex values of $k$ for every specified value of $\omega$ and $\phi$, thus characterizing a $k = k(\omega)$ method and allowing to obtain the complex band structure of the periodic unit cell.

In the case of purely elastic media, since $p''=0$ for each property ($\mu''$, $D''$, $\alpha''$, and $\beta''$), matrices $\mathbf{A}_2$, $\mathbf{A}_1$, and $\mathbf{A}_0$ become frequency-independent (see \ref{eigenproblem_matrices}), and the eigenproblem can be reduced to the generalized eigenvalue problem
\begin{equation} \label{eigen_wk}
 ( k^2 \mathbf{A}_2(\varphi) + k \mathbf{A}_1(\varphi) + \mathbf{A}_0(\varphi) ) \mathbf{V} = \omega^2 \mathbf{B} \mathbf{V} \, .
\end{equation}

This formulation is equivalent to the general procedure commonly found in literature \cite{poggetto2020widening}, which can be applied as a $\omega = \omega(k)$ approach for the purely elastic case. Although this method does not allow to obtain the complex band structure of the unit cell, it requires a considerably reduced computational effort, since it allows to obtain a representative dispersion curve by scanning the contour of the irreducible Brillouin zone (IBZ) \cite{maurin2018probability}. In the case of a square unit cell of side length $a$, the IBZ is defined as the triangular region delimited by the high-symmetry points $\Gamma$ ($k_x = k_y = 0$), X ($k_x=\pi/a, k_y=0$), and M ($k_x = k_y = \pi/a$).


\section{Results} \label{results}


For the numerical computations, two different materials are considered to create the hierarchical plate PCs using a hard (purely elastic behavior) and a soft phase (viscoelastic behavior), chosen as lead and rubber, respectively. Such materials were also chosen due to the large mismatch between their mechanical properties.
Lead properties are given by: shear storage modulus $\mu_1' = 14.9$ GPa, Poisson's ratio $\nu_1 = 0.3699$, and specific mass density $\rho_1 = 11600$ kg/m$^3$.
Rubber properties are given by: shear storage modulus $\mu_2' = 0.71$ MPa, Poisson's ratio $\nu_2 = 0.4998$, and specific mass density $\rho_2 = 1260$ kg/m$^3$ \cite{krushynska2016visco}. These materials are combined in two distinct configurations, here named as type A and type B. The type A configuration consists of a hard matrix (lead) with soft inclusions (rubber); the type B configuration consists of a soft matrix (rubber) with hard inclusions (lead). Increasing levels of shear viscosity are considered for the soft phase, using the values of $\mu''_2 = 5$, $10$, and $25$ Pa$\cdot$s \cite{krushynska2016visco}.

The results are presented considering the normalized frequency with respect to the matrix material for each configuration, given by $\overline{\omega}_i = \omega a/c_i$, where $c_i = \sqrt{\mu'_i/\rho_i}$ is the transverse wave speed (non-dispersive) corresponding to the $i$-th matrix material, thus yielding $c_1 = 1133$ m/s for lead (type A matrix) and $c_2 = 23.7$ m/s for rubber (type B matrix). For the maximum frequency $f_{\max} = 2$ kHz and thickness $h=3$ mm, the smallest wavelength of flexural waves for each medium are given by $\lambda_{\min} = \min( \sqrt[4]{D_1/\rho_1 h} \sqrt{2\pi/f_{\max}} , \sqrt[4]{D_2/\rho_2 h} \sqrt{2\pi/f_{\max}} ) = 11.4$ mm. We also select a square lattice of side $a = 100$ mm, which ensures $h << a$, thus indicating the applicability of plate theories. The normalized frequencies are shown for $\overline{\omega}_1 \in [0,1]$ (corresponding to $[0, 1.8]$ kHz) and $\overline{\omega}_2 \in [0,8]$ (corresponding to $[0, 302]$ Hz), to consider approximately the same number of propagating branches for each PC type.

The considered structures present increasing orders of hierarchy, and consequently, inclusion filling fractions (see Eq. (\ref{filling_fraction})), namely, no hierarchy ($\phi_0 = 0.1111$), first-order hierarchy ($\phi_1 = 0.2099$), and second-order hierarchy ($\phi_2 = 0.2977$).

A brief comparison between the band diagrams obtained using distinct plate theories is given in \ref{fe_plate_comparison} to justify the use of the Mindlin-Reissner plate theory.
A comparison between the results obtained using the FE and the PWE solutions with the $\omega = \omega(k)$ approach is given in \ref{validation_section}.
The next sections present results computed using the PWE methods. First, the main characteristics of the dispersion diagrams are analyzed for the purely elastic case using the $\omega = \omega(k)$ approach, which allows to investigate the contour of the IBZ. Finally, the effects of increasing viscosity levels and hierarchical orders are analyzed considering the dispersion diagrams obtained using the $k=k(\omega)$ formulations.

\subsection{Purely elastic case}

\begin{figure}[h!]
  \centering
  \makebox[\textwidth][c]{
  \includegraphics[width=15cm]{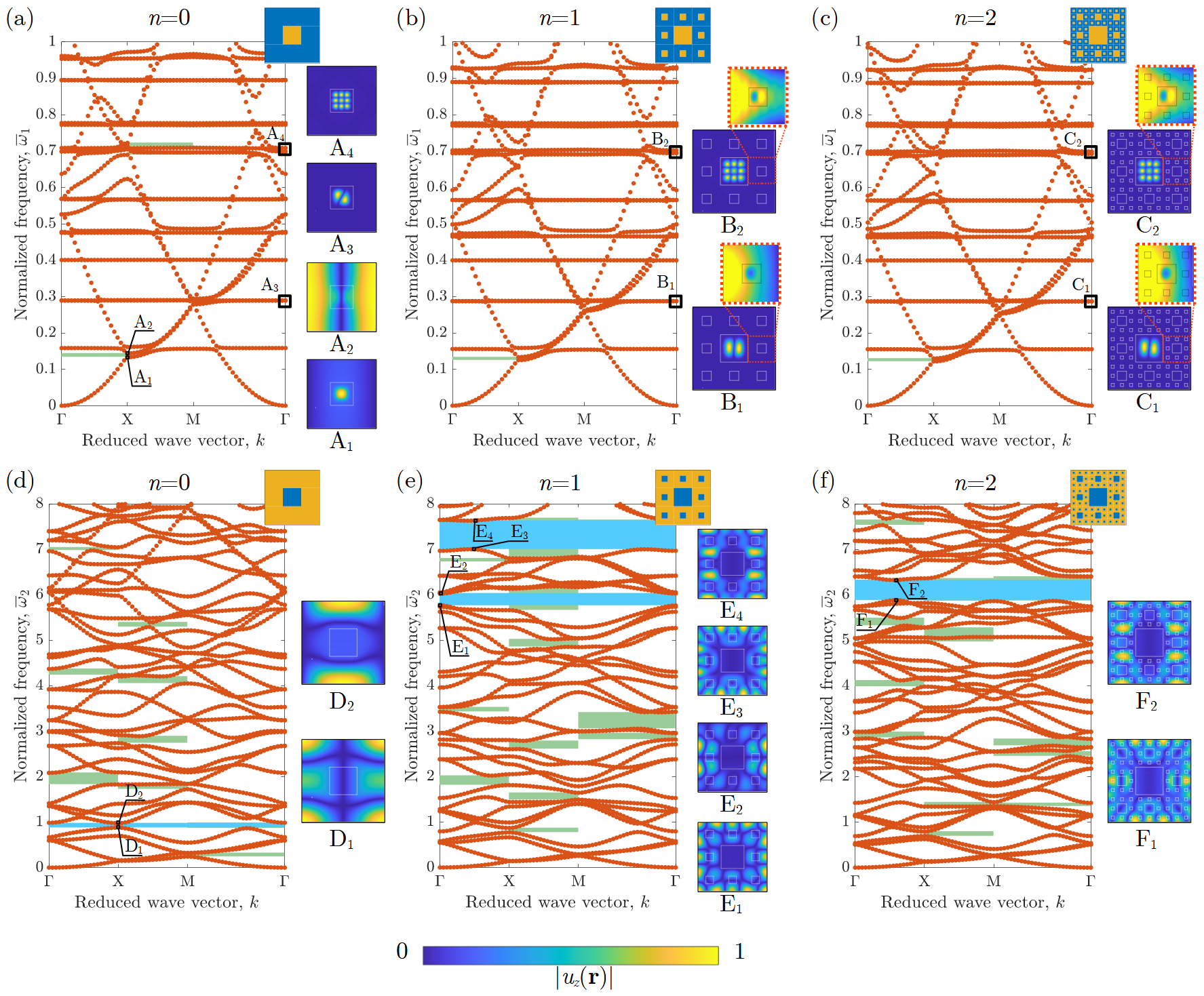}
  }
  \caption{
  Computed dispersion relations obtained   considering the type A PC (hard matrix, soft inclusions) for orders (a) $n=0$, (b) $n=1$, and (c) $n=2$, and type B PC (soft matrix, hard inclusions) for orders (d) $n=0$, (e) $n=1$, and (f) $n=2$.
  Wave modes of interest are indicated for each band diagram with the corresponding captions.
  The colorbar represents normalized out-of-plane displacements. Enlarged regions of the unit cells are also indicated with a modified color threshold.
  }
  \label{results_elastic}
\end{figure}

Let us initially consider the type A configuration (hard matrix and soft inclusions). Several flat bands (zero group velocity \cite{brillouin2013wave}, $\frac{\partial \omega}{\partial k}=0$) are computed for the hierarchical order $n=0$ (Figure \ref{results_elastic}a), with no full band gaps opened at their corresponding frequencies.
However, partial band gaps (i.e., for specific directions) are observed and highlighted in green. The partial band gap opened in the frequency range $\overline{\omega_1} \in [0.1334, 0.1438]$ is associated with wave modes A$_1$ and A$_2$.
Two wave modes are chosen to illustrate the flat branches, namely A$_3$ ($\overline{\omega}_1=0.2892$) and A$_4$ ($\overline{\omega}_1=0.7086$), representing wave modes with a strong energy concentration at the soft inclusion, thus indicating distinct locally resonant modes.

With an increase in the unit cell hierarchical order to $n=1$ (Figure \ref{results_elastic}b),
it is possible to notice that the frequencies of the represented wave modes, B$_1$ and B$_2$ (which correspond, respectively, to the previous modes A$_3$ and A$_4$) present a slight decrease, to $\overline{\omega}_1=0.2874$ and $\overline{\omega}_1=0.7020$, while the partial band gap is now located at $\overline{\omega_1} \in [0.1250, 0.1335]$
Also, although not noticeable using the normalized presented color scales, it is possible to verify an effect associated with the soft phase inclusions in this hierarchical order, which is highlighted by adjusting the color threshold and representing $1/9$ of the unit cell as the enlarged version of each wave mode. Thus, it is possible to notice, for modes B$_1$ and B$_2$, corresponding first-order and second-order resonant mode at the soft inclusions, although these effects are not significant when compared to the locally resonant mode at the central inclusion.

Finally, for the structure with $n=2$ (Figure \ref{results_elastic}c), the indicated wave modes, C$_1$ ($\overline{\omega}_1=0.2868$) and C$_2$ ($\overline{\omega}_1=0.6989$), respectively corresponding to A$_3$ (B$_1$) and A$_4$ (B$_2$), confirm the previously observed effect of the increase in the hierarchical order, i.e., flat bands present slightly decreased frequencies, while the partial band gap is now located at $\overline{\omega_1} \in [0.1215, 0.1298]$. No locally resonant mechanisms particularly associated with the smallest inclusions were noticed, even when considering higher frequency ranges (not shown here for the sake of brevity).

For the type B configuration (soft matrix and hard inclusions), the band diagram computed for the structure with $n=0$ (Figure \ref{results_elastic}d) shows a full band gap (i.e., for all wave vectors) highlighted in light blue, opened for $\overline{\omega}_2 \in [0.88, 0.98]$. This band gap is associated with the wave modes indicated as D$_1$ and D$_2$, both corresponding to the high-symmetry point X, showing that most part of energy is concentrated at the matrix material, thus indicating the formation of a Bragg-type band gap.
Many other partial band gaps are highlighted in green.

For an increased hierarchical order of $n=1$ (Figure \ref{results_elastic}e), the band gap which was previously observed for $n=0$ is no longer present, and instead, new band gaps are formed in the frequency range $\overline{\omega}_2 \in [5.76, 6.03]$, associated with the high-symmetry point $\Gamma$, and $\overline{\omega}_2 \in [7.00, 7.63]$, associated with the wave vector $\mathbf{k} = 0.45 \pi/a \, \hat{\mathbf{i}}$. The wave modes associated with the edges of these band gaps, denoted respectively as E$_1$/E$_2$ and E$_3$/E$_4$, indicate a noticeable concentration of energy at the matrix material, where no scatterers are present.

These effects continue to be observed for the hierarchical order $n=2$ (Figure \ref{results_elastic}f), where a band gap is now opened in the frequency range $\overline{\omega}_2 \in [5.88, 6.33]$, with wave modes at the edges denoted as F$_1$ and F$_2$, respectively, associated with the wave vector $\mathbf{k} = 0.60 \pi/a \,  \hat{\mathbf{i}}$. Although wave mode F$_1$ is similar to wave mode E$_1$, wave mode F$_2$ presents a displacement profile of a higher order, with more points of energy maxima localized between the scatterers included in this hierarchical order.

\clearpage

\subsection{Complex band structures}

In this section, we present the complex band diagrams for both type A and type B PCs, obtained using the EPWE formulation (Eq. (\ref{poly_eigen_kw_matrix})).
Since Figure \ref{results_elastic} indicates that all wave modes associated with the edges of band gaps lie in the $\Gamma$X direction, we restrict the analysis to $\varphi=0$,
and show the corresponding results using diagrams with positive (negative) numbers corresponding to the real (imaginary) part of the complex $k$ values. As the real part of the wave vector is restricted to the first Brillouin zone, Re$(k)a/\pi$ is contained in the $[0,1]$ interval. For the imaginary part, Im$(k)a/\pi$ is arbitrarily shown in the $[0,2]$ interval. Wave modes corresponding to the branch with the smallest imaginary part (least attenuated waves) for frequencies of interest are computed using Eq. (\ref{disp_final}) and shown for three consecutive unit cells.

\subsubsection*{Type A PC}

\begin{figure}[h!]
  \centering
  \makebox[\textwidth][c]{
  \includegraphics[width=15cm]{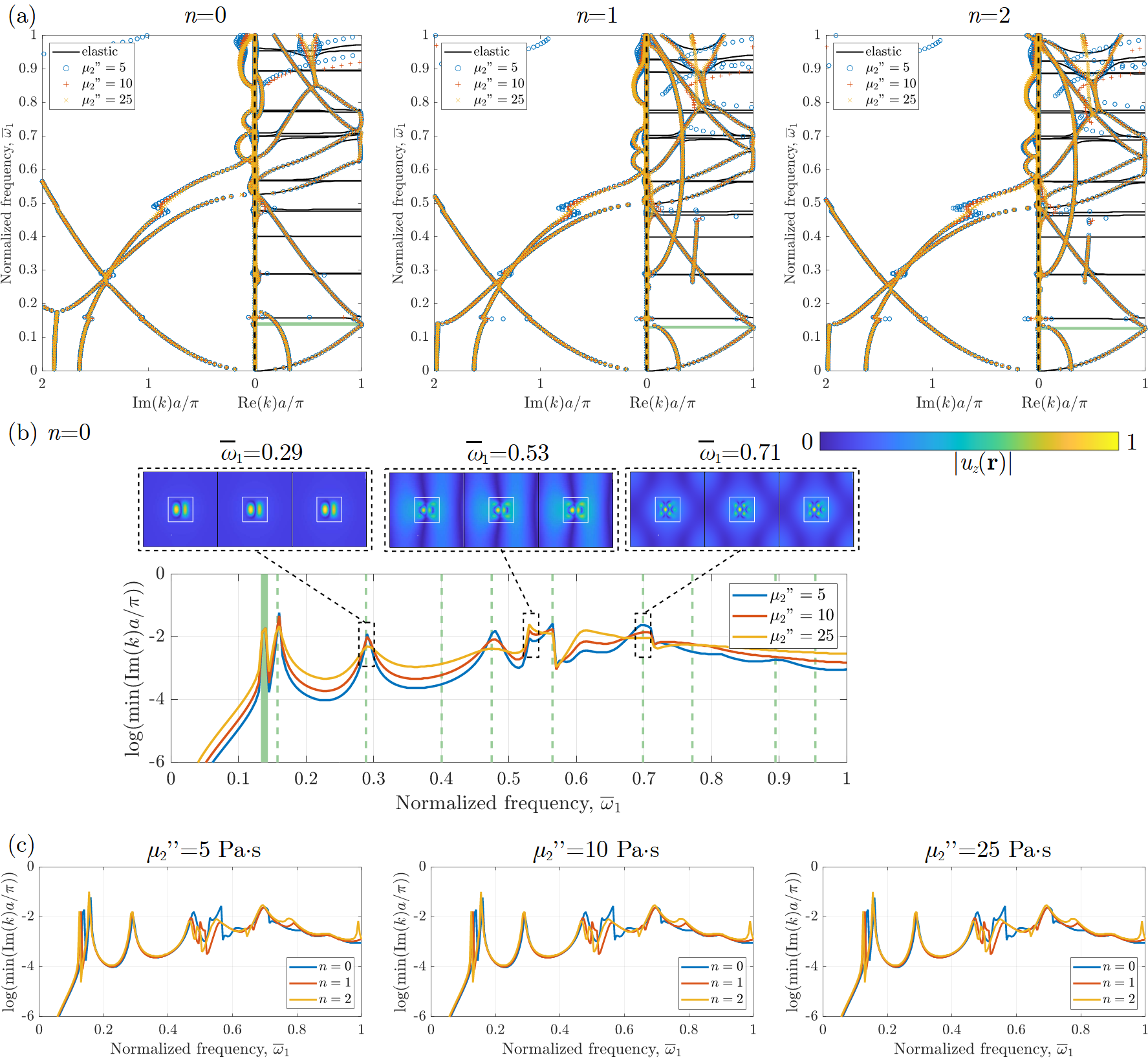}
  }
  \caption{
  Complex band structures for the type A PC considering (a) increasing shear viscosity levels for the inclusions ($\mu_2''$) and increasing orders of hierarchy ($n$).
  (b) Smallest attenuation for the $\Gamma$X direction considering the first hierarchical order ($n=0$) with increasing shear viscosity levels.
  (c) Comparison between smallest attenuation for increasing hierarchical orders at fixed shear viscosity levels.
  Green regions denote partial band gaps and vertical dashed lines indicate the frequencies of flat branches. The colorbar represents normalized out-of-plane displacements.
  }
  \label{complex_band_diagram_typea}
\end{figure}

We begin by showing the results for both the purely elastic cases, computed with the $\omega=\omega(k)$ approach (Eq. (\ref{eigen_wk})), as presented in last section (and thus yielding only real branches), and the complex band diagrams, computed with the $k = k(\omega)$ approach (Eq. (\ref{poly_eigen_kw_matrix})), for increasing hierarchical orders $n$ (Figure \ref{complex_band_diagram_typea}a). The $k=k(\omega)$ approach yields more bands than the $\omega = \omega(k)$ approach, which is not able to fully describe the complex band structure due to the assumption of real wave vectors \cite{laude2009evanescent,andreassen2013analysis}.
It is also interesting to notice that, although the $k=k(\omega)$ approach fails to match the $\omega = \omega(k)$ results for the real part of the flat branches (see \ref{validation_section}), the corresponding imaginary part of such branches demonstrates evidence of the presence of flat branches in the form of swift changes in its derivative ($\frac{\partial \omega}{\partial k}$), with the most noticeable example occurring near $\overline{\omega}_1=0.5$ for all hierarchical orders. Although the real part of the computed branches presents noticeable variations between $n=0$ and $n=1$, the same cannot be said between $n=1$ and $n=2$, which reinforces the observations made regarding Figures \ref{results_elastic}a--c. The same can be said with respect to the imaginary part of the computed branches, which present slight variations close to $\overline{\omega}_1=0.5$.

Figure \ref{complex_band_diagram_typea}b presents the minimum of the imaginary part of the computed wave vectors for $n=0$, comparing distinct viscosity levels $\mu_2''$, showing also the wave modes corresponding to the least attenuated wave for $\mu_2''=25$ Pa$\cdot$s.
The highlighted green region indicates the previously computed partial band gap (see Figure \ref{results_elastic}a), while the vertical dashed lines indicate flat branches.
The wave modes corresponding to the peaks shown at $\overline{\omega}_1=0.29$ and $\overline{\omega}_1=0.71$ correspond to the wave modes labeled as C$_1$ and C$_2$ in Figure \ref{results_elastic}c. In the vicinity of these frequencies, an increase in the viscosity level of the inclusions imply in a decrease in the peak attenuation level, which is a well known property when considering simple locally resonant systems \cite{inman1994engineering}. On the other hand, the attenuation levels are increased upon an increase in viscosity level for frequency regions between peaks corresponding to flat branches, while the attenuation associated with the partial band gap remains constant. This observation is also true for regions which present local peaks which are not associated with flat branches, as in the wave mode illustrated at $\overline{\omega}_1=0.53$, which lies between the flat branches localized at $\overline{\omega}_1=0.46$ and $\overline{\omega}_1=0.56$. Thus, one may conclude that an increase in the viscosity level of the inclusions is detrimental for the frequencies associated with flat branches, but beneficial for frequencies comprised between flat branches.

The influence of increasing orders of hierarchy for fixed viscosity levels on the smallest attenuated waves is highlighted in Figure \ref{complex_band_diagram_typea}c. The most noticeable effect is the shifting of peaks to lower frequencies as the structural hierarchy order is increased, which is most easily noticeable in the range $\overline{\omega}_1 \in [0.4, \, 0.6]$.
From the presented results, it becomes clear that an increase in the hierarchical order of this type of structure does not lead to an improvement in the resulting levels of attenuation.
However, due to the increasing filling fractions (see Eq. (\ref{filling_fraction})) of inclusions with lower densities when compared to the matrix material, the specific mass densities of these configurations, computed as $\overline{\rho}_0^{(\text{A})} = 10444$ kg/m$^3$, $\overline{\rho}_1^{(\text{A})} = 9417$ kg/m$^3$, and $\overline{\rho}_2^{(\text{A})} = 8504$ kg/m$^3$ for the type A PC, indicate that it is possible to significantly decrease the resulting structure weight while practically keeping the same attenuation levels.

\clearpage

\subsubsection*{Type B PC}

\begin{figure}[h!]
  \centering
  \makebox[\textwidth][c]{
  \includegraphics[width=12cm]{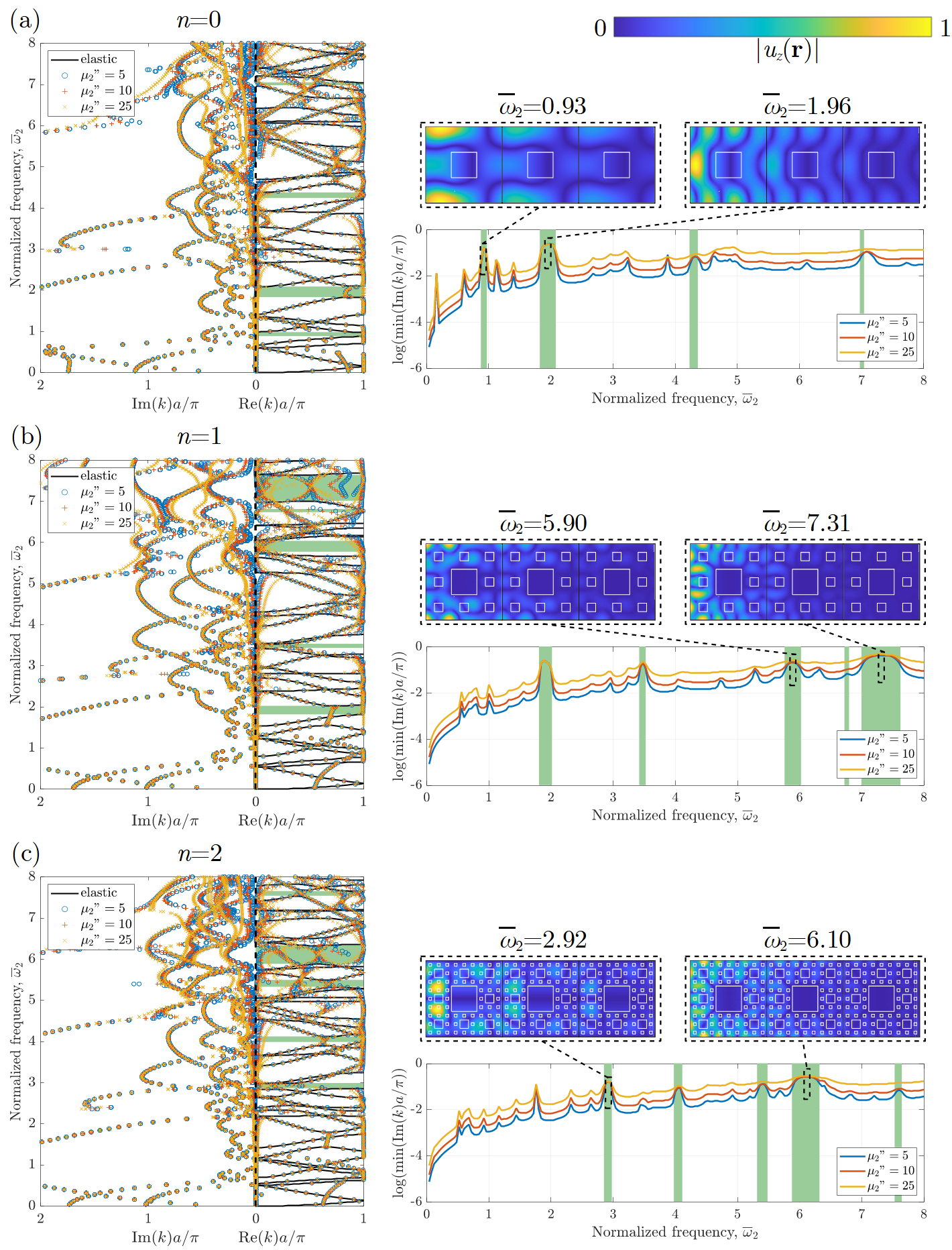}
  }
  \caption{
  Complex band structures for the type B PC and wave modes corresponding to the least attenuated branches at the indicated normalized frequencies for the hierarchical orders (a) $n=0$, (b) $n=1$, and (c) $n=2$, with increasing levels of matrix material shear viscosity ($\mu_2''$).
  Green regions denote partial band gaps.
  The colorbar represents normalized out-of-plane displacements.
  }
  \label{complex_band_diagram_typeb}
\end{figure}

In a similar manner, the complex band diagrams obtained for the type B PC are shown in Figure \ref{complex_band_diagram_typeb}, with partial band gaps in the $\Gamma$X direction highlighted in green. The shown wave modes are computed at the central frequency of partial band gaps. Some dispersion branches may be found even within frequency ranges considered as band gaps by the $\omega = \omega(k)$ approach (i.e., with no real part), which implies that, when using a $k=k(\omega)$ approach, band gaps must be defined as frequency ranges where the imaginary part of all waves present a non-zero value. This interpretation yields a more robust definition of band gap, which can be regarded as a viable metric for many applications, including structural optimization \cite{ribeiro2022robust}.

For the structure with hierarchical order $n=0$ (Figure \ref{complex_band_diagram_typeb}a),
the imaginary part of the complex band diagram presents increasingly differences for $\overline{\omega}_2 > 5$. More interestingly, the least attenuated waves computed for the $\Gamma$X direction present a monotonic non-decreasing behavior with the increase of viscosity levels. For the partial band gaps with central frequencies indicated as $\overline{\omega}_2 = 0.93$ (which is also part of a full band gap) and $\overline{\omega}_2 = 1.96$, the increase in viscosity levels do not lead to a decrease in attenuation, keeping a constant level of attenuation, thus presenting the opposite behavior as the type A PC. For frequency ranges between partial band gaps, the levels of attenuation are typically increased. The wave modes associated with the indicated frequencies show the formation of band gaps due to Bragg scattering, as previously stated regarding Figure \ref{results_elastic}.

Similar observations can be made with respect to the hierarchical orders $n=1$ (Figure \ref{complex_band_diagram_typeb}b) and $n=2$ (Figure \ref{complex_band_diagram_typeb}c). 
For the hierarchical order $n=1$ ($n=2$), the imaginary part of the complex band diagram presents increasing differences for $\overline{\omega}_2 > 3$ ($>2$). The same monotonic behavior is observed for the least attenuated waves for increasing levels of viscosity, with the wave modes associated with partial band gaps shown at frequencies $\overline{\omega}_2 = 5.90$ and $\overline{\omega}_2 = 7.31$ ($\overline{\omega}_2 = 2.92$ and $\overline{\omega}_2 = 6.10$).

In the case of the type B structure, the equivalent specific mass densities for the type B PCs are given, for increasing hierarchical orders, as $\overline{\rho}_0^{(\text{B})} = 2356$ kg/m$^3$, $\overline{\rho}_1^{(\text{B})} = 3383$ kg/m$^3$, and $\overline{\rho}_2^{(\text{B})} = 4296$ kg/m$^3$. Thus, although increasing the structure hierarchical order may be useful to manipulate the band gap distribution, this also implies in an increase in the overall mass density of the unit cell.

\clearpage
\section{Conclusions} \label{conclusions}

In conclusion, we have presented a PWE formulation for the computation of the complex band structure of plate PCs using the Mindlin-Reissner plate theory and the Kelvin-Voigt model to represent viscoelastic effects, which were used to investigate the influence of increasing levels of viscosity and hierarchical structuring. The proposed PWE method presents a very good agreement with the FE-based method with a considerable reduction in the dimension of the problem.

For the PC configuration of a hard purely elastic matrix with soft viscoelastic inclusions (type A), many flat bands are computed, and although no considerable band gaps are opened due to these flat bands, they are associated with localized increases in attenuation when considering the least attenuated waves for the complex wave vector in a given direction. Also, due to the locally resonant mechanisms associated with these flat bands, an increase in the viscosity level of the soft inclusions leads to a decrease in the attenuation at these peaks and an increase in the attenuation in the frequency ranges between peaks. Although an increase in the structure hierarchical order does not lead to significant changes in the unit cell attenuation, it can be used to reduce the unit cell specific mass density.

For the PC constituted by a soft viscoelastic matrix with hard purely elastic inclusions (type B), Bragg scattering band gaps are opened for every considered hierarchical order (not necessarily preserved between consecutive hierarchical orders), with wave modes indicating a distributed energy profile at the matrix material. Although increasing the hierarchical order may be used to tune the band gaps location, it also implies in an increase in the unit cell specific mass density. For this type of PC, an increase in the viscosity levels of the soft phase does not hinder the attenuation of waves inside partial band gaps, while monotonically enhancing (non-decreasing) the attenuation in frequency regions outside of band gaps.

Overall, for the type A PC, hierarchical structuring can be harnessed as a strategy of mass reduction. If the type B PC is preferred, hierarchical structuring can be used to manipulate the opening of band gaps at distinct frequency ranges at the cost of an increase in mass.

\section*{Acknowledgments}

VFDP and NMP are supported by the EU H2020 FET Open ``Boheme'' grant No. 863179.
EJPM and JMCDS thank the Brazilian funding agencies
CAPES (finance Code 001), 
CNPq (grants 313620/2018, 151311/2020-0 and 403234/2021-2),
FAPEMA (grants 02558/21, 02559/21 and 00680/22), and
FAPESP (grant 2018/15894-0).


\appendix

\section{Eigenproblem matrices} \label{eigenproblem_matrices}

After truncating the total number of reciprocal lattice vectors in Eqs. (\ref{mindlin_pwe_uz})--(\ref{mindlin_pwe_psiy}) for a total of $n_G$ plane waves, these equations can be written in the form of a polynomial eigenvalue problem as
\begin{equation} \label{poly_eigen_app}
 ( k^2 \mathbf{A}_2 + k \mathbf{A}_1 + \mathbf{A}_0 - \omega^2 \mathbf{B} ) \mathbf{V} = \mathbf{0} \, ,
\end{equation}
where matrices $\mathbf{A}_2$, $\mathbf{A}_1$, $\mathbf{A}_0$, and $\mathbf{B}$ can be partitioned as
\begin{equation}
\begin{aligned}
 &\mathbf{A}_2 = 
 \left[
 \begin{array}{ccc}
  \mathbf{A}_{2zz} & \mathbf{0} & \mathbf{0} \\
  \mathbf{0} & \mathbf{A}_{2xx} & \mathbf{A}_{2xy} \\
  \mathbf{0} & \mathbf{A}_{2yx} & \mathbf{A}_{2yy} \\
 \end{array}
 \right] \, ,
 \mathbf{A}_1 =
 \left[
 \begin{array}{ccc}
  \mathbf{A}_{1zz} & \mathbf{A}_{1zx} & \mathbf{A}_{1zy} \\
  \mathbf{A}_{1xz} & \mathbf{A}_{1xx} & \mathbf{A}_{1xy} \\
  \mathbf{A}_{1yz} & \mathbf{A}_{1yx} & \mathbf{A}_{1yy} \\
 \end{array}
 \right] \, , \nonumber\\
 &\mathbf{A}_0 =
 \left[
 \begin{array}{ccc}
  \mathbf{A}_{0zz} & \mathbf{A}_{0zx} & \mathbf{A}_{0zy} \\
  \mathbf{A}_{0xz} & \mathbf{A}_{0xx} & \mathbf{A}_{0xy} \\
  \mathbf{A}_{0yz} & \mathbf{A}_{0yx} & \mathbf{A}_{0yy} \\
 \end{array}
 \right] \, ,
 \mathbf{B} =
 \left[
 \begin{array}{ccc}
  \mathbf{B}_{zz} & \mathbf{0} & \mathbf{0} \\
  \mathbf{0} & \mathbf{B}_{xx} & \mathbf{0} \\
  \mathbf{0} & \mathbf{0} & \mathbf{B}_{yy} \\
 \end{array}
 \right] \, ,
\end{aligned}
\end{equation}
and the eigenvector $\mathbf{V}$ represents
\begin{equation}
 \mathbf{V} = \left\{
 \begin{array}{ccccccccc}
  \hat{u}_z (\mathbf{G}_1) & \ldots & \hat{u}_z (\mathbf{G}_{n_G}) &
  \hat{\psi}_x (\mathbf{G}_1) & \ldots & \hat{\psi}_x (\mathbf{G}_{n_G}) &
  \hat{\psi}_y (\mathbf{G}_1) & \ldots & \hat{\psi}_y (\mathbf{G}_{n_G}) 
 \end{array}
 \right\}^T \, .
\end{equation}

The terms of the matrices that form $\mathbf{A}_2$ are given by
\begin{equation}
\begin{aligned}
 [\mathbf{A}_{2zz}]_{ij} &= \kappa (\hat{\mu}'_{ij}-\text{i}\omega\hat{\mu}''_{ij}) h \, , \\
 [\mathbf{A}_{2xx}]_{ij} &= (\hat{D}'_{ij}-\text{i}\omega\hat{D}''_{ij}) \cos^2 \varphi + (\hat{\beta}'_{ij}-\text{i}\omega\hat{\beta}''_{ij}) \sin^2 \varphi \, , \\
 [\mathbf{A}_{2yy}]_{ij} &= (\hat{D}'_{ij}-\text{i}\omega\hat{D}''_{ij}) \sin^2 \varphi + (\hat{\beta}'_{ij}-\text{i}\omega\hat{\beta}''_{ij}) \cos^2 \varphi \, , \\
 [\mathbf{A}_{2yx}]_{ij} = [\mathbf{A}_{2xy}]_{ij} &= (\hat{\alpha}'_{ij}-\text{i}\omega\hat{\alpha}''_{ij} + \hat{\beta}'_{ij}-\text{i}\omega\hat{\beta}''_{ij}) \sin \varphi \cos \varphi \, ; \\
\end{aligned}
\end{equation}
for the terms that form $\mathbf{A}_1$, one may write
\begin{equation}
\begin{aligned}
 [\mathbf{A}_{1zz}]_{ij} &= \kappa (\hat{\mu}'_{ij}-\text{i}\omega\hat{\mu}''_{ij}) h ( \cos \varphi (G_{xi}+G_{xj}) + \sin \varphi (G_{yi}+G_{yj}) ) \, , \\
 [\mathbf{A}_{1xx}]_{ij} &= (\hat{D}'_{ij}-\text{i}\omega\hat{D}''_{ij}) \cos \varphi (G_{xi}+G_{xj}) + (\hat{\beta}'_{ij}-\text{i}\omega\hat{\beta}''_{ij}) \sin \varphi (G_{yi}+G_{yj}) \, ,\\
 [\mathbf{A}_{1yy}]_{ij} &= (\hat{D}'_{ij}-\text{i}\omega\hat{D}''_{ij}) \sin \varphi (G_{yi}+G_{yj}) + (\hat{\beta}'_{ij}-\text{i}\omega\hat{\beta}''_{ij}) \cos \varphi (G_{xi}+G_{xj}) \, , \\
 [\mathbf{A}_{1zx}]_{ij} = -[\mathbf{A}_{1xz}]_{ij} &= \text{i} \kappa (\hat{\mu}'_{ij}-\text{i}\omega\hat{\mu}''_{ij}) h \cos \varphi \, , \\
 [\mathbf{A}_{1zy}]_{ij} = -[\mathbf{A}_{1yz}]_{ij} &= \text{i} \kappa (\hat{\mu}'_{ij}-\text{i}\omega\hat{\mu}''_{ij}) h \sin \varphi \, , \\
 [\mathbf{A}_{1xy}]_{ij} = [\mathbf{A}_{1yx}]_{ji} &= (\hat{\alpha}'_{ij}-\text{i}\omega\hat{\alpha}''_{ij}) (\sin \varphi \, G_{xi}+ \cos \varphi \, G_{yj}) + (\hat{\beta}'_{ij}-\text{i}\omega\hat{\beta}''_{ij}) (\cos \varphi \, G_{yi}+ \sin \varphi \, G_{xj}) \, ; \\
\end{aligned}
\end{equation}
while the submatrices of $\mathbf{A}_0$ are given by
\begin{equation}
\begin{aligned}
 [\mathbf{A}_{0zz}]_{ij} &= \kappa (\hat{\mu}'_{ij}-\text{i}\omega\hat{\mu}''_{ij}) h ( G_{xi}G_{xj}+G_{yi}G_{yj} ) \, , \\
 [\mathbf{A}_{0xx}]_{ij} &= (\hat{D}'_{ij}-\text{i}\omega\hat{D}''_{ij}) G_{xi} G_{xj} + (\hat{\beta}'_{ij}-\text{i}\omega\hat{\beta}''_{ij}) G_{yi} G_{yj} + \kappa (\hat{\mu}'_{ij}-\text{i}\omega\hat{\mu}''_{ij}) h \, ,\\
 [\mathbf{A}_{0yy}]_{ij} &= (\hat{D}'_{ij}-\text{i}\omega\hat{D}''_{ij}) G_{yi} G_{yj} + (\hat{\beta}'_{ij}-\text{i}\omega\hat{\beta}''_{ij}) G_{xi} G_{xj} + \kappa (\hat{\mu}'_{ij}-\text{i}\omega\hat{\mu}''_{ij}) h \, , \\
 [\mathbf{A}_{0zx}]_{ij} = - [\mathbf{A}_{0xz}]_{ji} &= \text{i} \kappa (\hat{\mu}'_{ij}-\text{i}\omega\hat{\mu}''_{ij}) h G_{xi} \, , \\
 [\mathbf{A}_{0zy}]_{ij} = - [\mathbf{A}_{0yz}]_{ji} &= \text{i} \kappa (\hat{\mu}'_{ij}-\text{i}\omega\hat{\mu}''_{ij}) h G_{yi} \, , \\
 [\mathbf{A}_{0xy}]_{ij} = [\mathbf{A}_{0yx}]_{ji} &= (\hat{\alpha}'_{ij}-\text{i}\omega\hat{\alpha}''_{ij}) G_{xi} G_{yj} + (\hat{\beta}'_{ij}-\text{i}\omega\hat{\beta}''_{ij}) G_{yi} G_{xj} \, ; \\
\end{aligned}
\end{equation}
and finally, matrix $\mathbf{B}$ is partitioned as
\begin{equation}
\begin{aligned}
 [\mathbf{B}_{zz}]_{ij} &= \hat{\rho}_{ij} h \, , \\
 [\mathbf{B}_{xx}]_{ij} = [\mathbf{B}_{yy}]_{ij} &= \hat{\rho}_{ij} h^3/12 \, .\\
\end{aligned}
\end{equation}

The solution of Eq. (\ref{poly_eigen_app}) can be performed considering both a $k=k(\omega)$ approach, for the viscoelastic case (EPWE method), or a $\omega = \omega(k)$ approach, for the purely elastic case (PWE method).

\section{Band diagrams obtained using distinct plate theories} \label{fe_plate_comparison}
\setcounter{figure}{0}

In this section, we compare the band diagrams obtained using the FE method considering both the Kirchhoff and Mindlin plate theories. The band diagrams are computed considering elastic structures, with stiffness and mass matrices obtained for Kirchhoff and Mindlin plate elements \cite{cook2001concepts}. Periodic boundary conditions are enforced \cite{mace2008modelling} using the implementation as given in \cite{poggetto2021band}. Due to the symmetry of the unit cells, we restrict our analysis to the contour of the first Brillouin zone \cite{maurin2018probability}, obtaining the values $\omega = \omega(k)$ of propagating frequencies. The number of elements used in the FE models is increased until no significant changes are observed in the dispersion diagrams, which is achieved with $6,561$ elements, corresponding to $20,172$ degrees-of-freedom for both the Kirchhoff (C$_1$ continuity) and Mindlin plate elements (C$_0$ continuity), yielding eigenproblems with an associated dimension of $19,683$.

\begin{figure}[H]
  \centering
  \makebox[\textwidth][c]{
  \includegraphics[width=15cm]{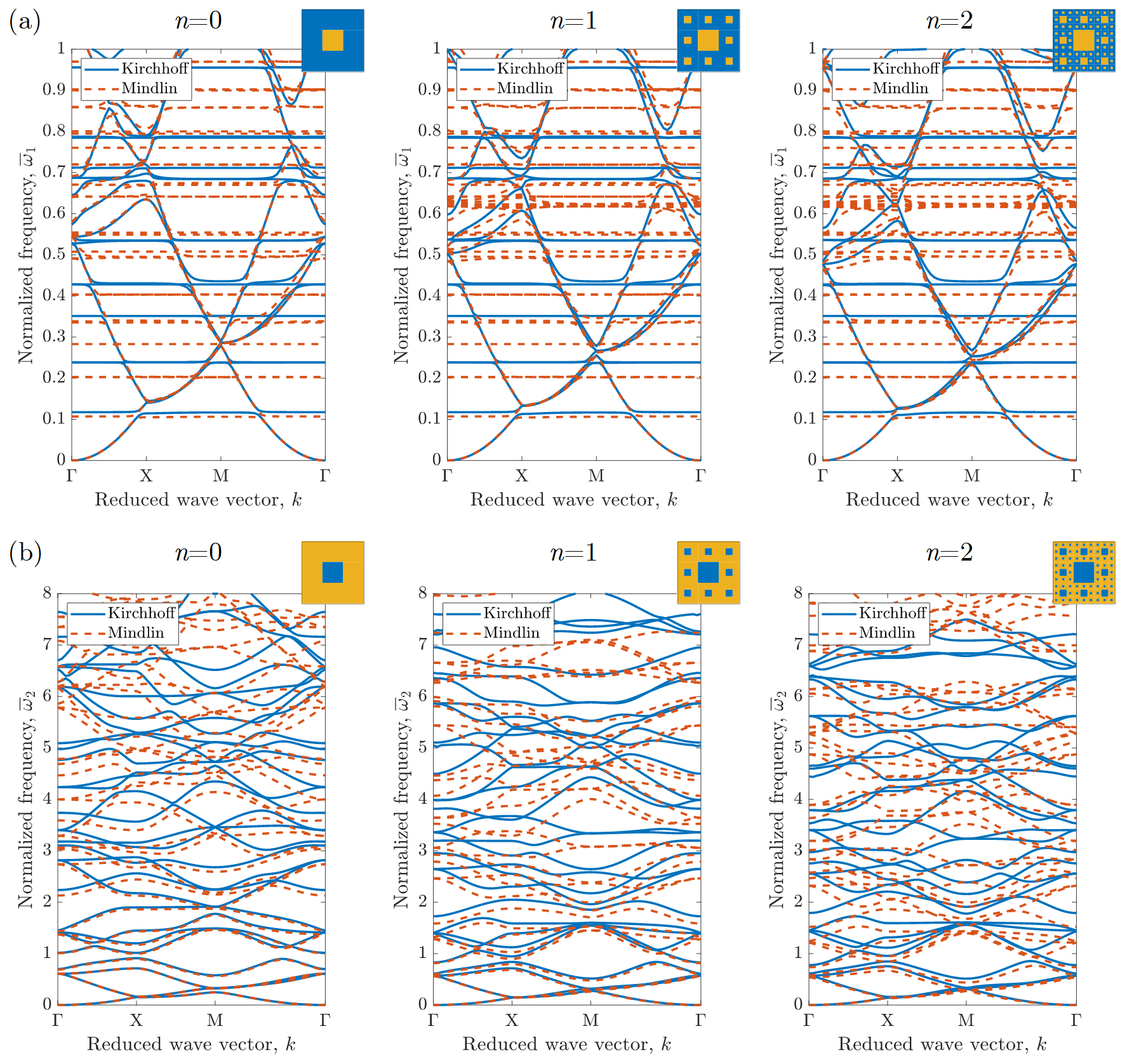}
  }
  \caption{
  Comparison between the band diagrams computed for the $\omega=\omega(k)$ formulation using the FE method considering the Kirchhoff and Mindlin plate formulations.
  Results are shown for the (a) type A and (b) type B configurations, with increasing hierarchical orders ($n=0$, $1$, and $2$).
  }
  \label{fe_comparison_kirchhoff_vs_mindlin}
\end{figure}

Initially considering the type A configuration (hard matrix and soft inclusions, Figure \ref{fe_comparison_kirchhoff_vs_mindlin}a), it is possible to notice that the differences between the results obtained considering each plate theory are especially related with the flat bands for the structure with $n=0$. Also, as the order of hierarchy increases to $n=1$ and $n=2$, it is possible to notice that a large number of branches is revealed by the Mindlin model but not by the Kirchhoff model, especially for $\overline{\omega}_1 > 0.5$. The differences in the branches with non-zero group velocities \cite{brillouin2013wave}, however, are not significant.

For the case of the type B configuration (soft matrix and hard inclusions, Figure \ref{fe_comparison_kirchhoff_vs_mindlin}b), no flat branches are computed considering either plate theory. On the other hand, the differences between the obtained band diagrams become noticeable in the frequency ranges $\overline{\omega}_2 > 3$ for $n=0$, $\overline{\omega}_2 > 2$ for $n=1$, and $\overline{\omega}_2 > 1$ for $n=2$. These results indicate that not only the Mindlin plate theory becomes necessary for higher frequencies, but also for configurations of higher fractal order. Thus the Mindlin plate theory should be preferred for both configurations.

\section{Validation of the PWE method} \label{validation_section}
\setcounter{figure}{0}

In this section, we compare the results obtained using the $\omega = \omega(k)$ formulation computed using the FE and PWE methods (Eq. (\ref{eigen_wk})).
The band diagrams are not analyzed with respect to band gap formation and associated wave modes, restricting ourselves to the consideration of convergence between different methods.
For the FE method, band diagrams are computed considering purely elastic materials using Mindlin plate elements, following the same procedure as presented in \ref{fe_plate_comparison}, thus leading to eigenproblems with an associated dimension of $19,683$.
The PWE results are computed with an increasing number of plane waves for both the conventional (PWE) and improved (IPWE) formulations, reaching a total of $961$ plane waves, with eigenproblems of dimension $2,883$, thus representing a reduction of around $85\%$ with respect to the FE formulation.

\begin{figure}[H]
  \centering
  \makebox[\textwidth][c]{
  \includegraphics[width=15cm]{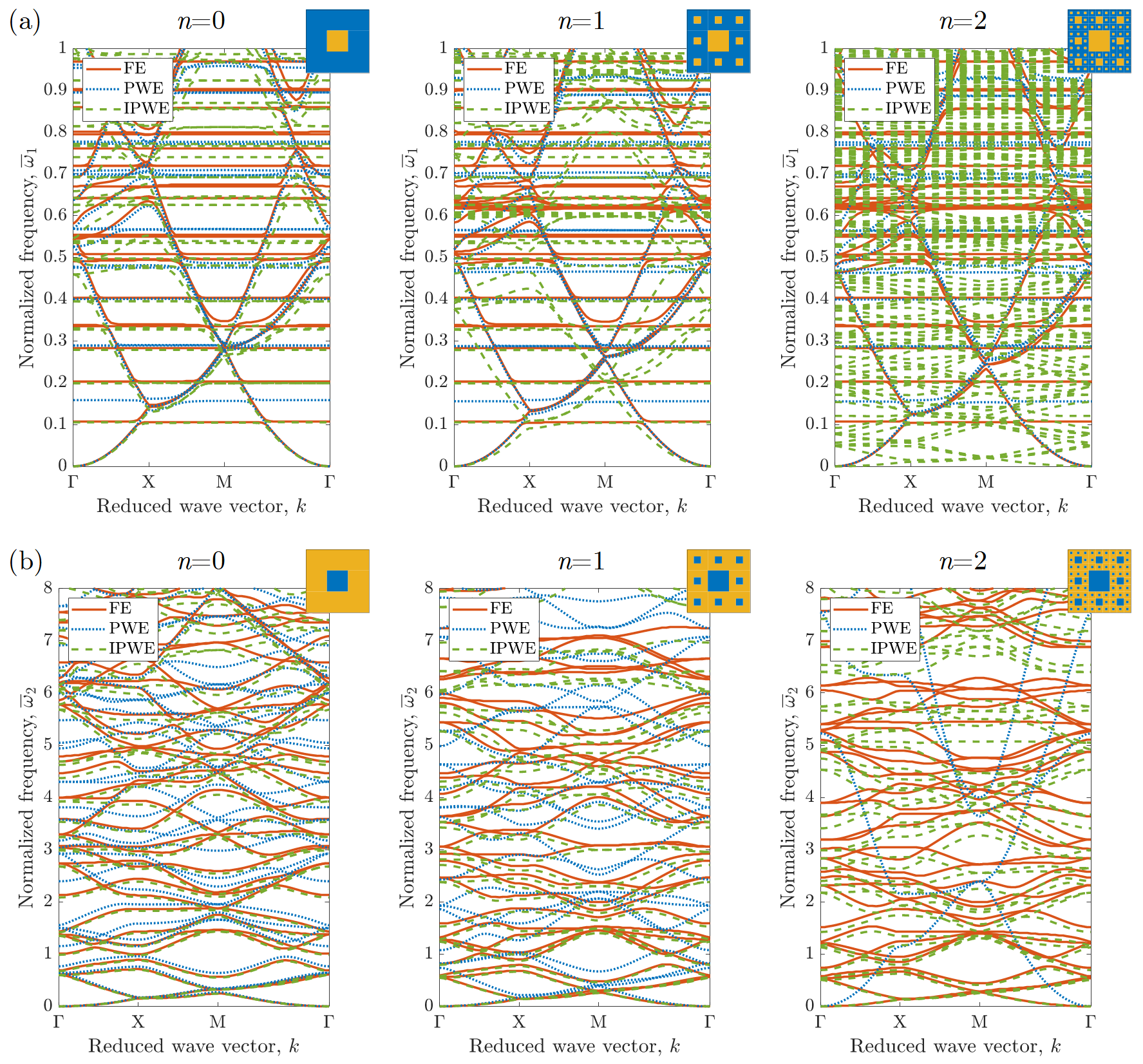}
  }
  \caption{
  Comparison between the band diagrams computed for the $\omega=\omega(k)$ formulation using the FE, PWE, and IPWE methods. 
  Results are shown for the (a) type A and (b) type B configurations, with increasing hierarchical orders ($n=0$, $1$, and $2$).
  }
  \label{validation_pwe_vs_fe_mindlin}
\end{figure}

For the type A configuration (hard matrix and soft inclusions, Figure \ref{validation_pwe_vs_fe_mindlin}a), the band diagram computed for the structure with $n=0$ shows an excellent agreement between the branches with non-zero group velocity (derivative of each branch with respect to the wavenumber, $\frac{\partial \omega}{\partial k}$ \cite{brillouin2013wave}) computed using the FE and the PWE, with a slight underestimation of the propagating frequencies by the IPWE. Several flat bands (zero group velocity, $\frac{\partial \omega}{\partial k}=0$) are computed  using the FE method, with no associated band gaps. A disagreement is noticed for the frequencies of such flat bands computed using the PWE, while the IPWE presents a considerably better agreement. With an increase in the unit cell hierarchical order to $n=1$, it is possible to notice a very good agreement between the results obtained with the PWE and the FE formulations, while the frequencies of flat bands are incorrectly estimated. In this case, the IPWE underestimates the branches with non-zero group velocity, while correctly estimating the frequencies of the flat bands. Finally, for the structure with $n=2$, the results obtained by the PWE maintain the previous trend, presenting a very good agreement for the non-zero group velocity branches and poor agreement for the flat bands. At this hierarchical order, the results obtained with the IPWE method show a poor agreement with the FE-based results, largely underestimating the propagating frequencies. It is important to notice that the computation of flat bands is possible when considering a $\omega = \omega(k)$ method; however, if one considers the $k = k(\omega)$ approach for the computation of the complex band structure, these flat bands will only be revealed if their exact frequencies are considered in the computation, thus not necessarily always being computed.

For the type B configuration (soft matrix and hard inclusions, Figure \ref{validation_pwe_vs_fe_mindlin}b), the band diagram computed for the structure with $n=0$ shows an excellent agreement between the FE and the IPWE, while the PWE method overestimates the propagating frequencies. For an increased hierarchical order of $n=1$, the IPWE method still presents a good agreement with the FE-based results, while the PWE considerably overestimates the propagating frequencies. Finally, for the hierarchical order $n=2$, only the IPWE yields results with a reasonable agreement with the ones obtained using the FE method, although slightly underestimating the propagating frequencies.

It is interesting to note that the PWE presents a better correlation with the FE for the type A PC, while the IPWE presents a better correlation for the type B PC. For this reason, the PWE is considered for the computations involving the type A PC, while the IPWE is used for the type B PC.

\bibliographystyle{elsarticle-num}
\bibliography{hierarchical_evanescent_plate}

\end{document}